\documentclass[runningheads]{llncs}

\usepackage{cite}
\usepackage[colorlinks=true,
    linkcolor=red,
    urlcolor=blue,citecolor=blue]{hyperref}
\usepackage{amsmath}
\usepackage{multirow}
\usepackage{algorithmic}
\usepackage{array}
\usepackage{amssymb}
\usepackage{booktabs}
\usepackage{stfloats}
\usepackage{url}
\usepackage{color}
\usepackage{xspace}
\usepackage{bm}

\usepackage[T1]{fontenc}
\usepackage{graphicx}
\usepackage[misc]{ifsym}

\newcommand{\etal}{\textit{et al}.\xspace}
\newcommand{\ie}{\textit{i}.\textit{e}.\xspace}
\newcommand{\eg}{\textit{e}.\textit{g}.\xspace}

\newcommand{\std}[1]{\tiny #1}
\newcommand{\mat}[1]{\boldsymbol{#1}}

\newcommand{\lossname}{LEDA\xspace}

\newlength\savewidth\newcommand\shline{\noalign{\global\savewidth\arrayrulewidth
  \global\arrayrulewidth 1.5pt}\hline\noalign{\global\arrayrulewidth\savewidth}}
  
\begin{document}
\title{Low-dose CT Denoising with Language-engaged Dual-space Alignment}

\titlerunning{LEDA for Low-dose CT Denoising}

\author{*}
\author{Zhihao Chen\inst{1} \and
Tao Chen\inst{1} \and Chenhui Wang\inst{1} \and Chuang Niu\inst{2} \and Ge Wang\inst{2}\textsuperscript{(\Letter)} \and
Hongming Shan\inst{1}\textsuperscript{(\Letter)}}

\authorrunning{Z. Chen et al.}

\institute{Institute of Science and Technology for Brain-Inspired Intelligence, Fudan University, Shanghai, China \\
\and
Center for Biotechnology and Interdisciplinary Studies, Rensselaer Polytechnic Institute, Troy, USA\\
\email{wangg6@rpi.edu, hmshan@fudan.edu.cn}
}

\maketitle

\begin{abstract}
While various deep learning methods were proposed for low-dose computed tomography (CT) denoising, they often suffer from over-smoothing, blurring, and lack of explainability. 
To alleviate these issues, we propose a plug-and-play \textbf{L}anguage-\textbf{E}ngaged \textbf{D}ual-space \textbf{A}lignment loss (\lossname) to optimize low-dose CT denoising models.
Our idea is to leverage large language models (LLMs) to align denoised CT and normal dose CT images in both the \emph{continuous perceptual space} and \emph{discrete semantic space}, which is the first LLM-based scheme for low-dose CT denoising.
\lossname involves two steps: the first is to pretrain an LLM-guided CT autoencoder, which can encode a CT image into continuous high-level features and quantize them into a token space to produce semantic tokens derived from the LLM's vocabulary; and the second
is to minimize the discrepancy between the denoised CT images and normal dose CT in terms of both encoded high-level features and quantized token embeddings derived by the LLM-guided CT autoencoder.
Extensive experimental results on two public LDCT denoising datasets demonstrate that our \lossname can enhance existing denoising models in terms of quantitative metrics and qualitative evaluation, and also provide explainability through language-level image understanding. Source code is available at \url{https://github.com/hao1635/LEDA}.
\end{abstract}
\section{Introdcution}

Over the past several years, various deep learning methods were developed with promising results for low-dose computed tomography (LDCT) denoising~\cite{wang2018image,redcnn,shan2019competitive,shan20183,wgan-vgg,gholizadeh2020deep, huang2021gan,corediff,litformer,lei2023ct, niu2022noise, jing2023inter, wang2023ctformer, wang2023masked,shen2022mlf,ge2022ddpnet,ge2023jccs}. 
Most of these methods rely on LDCT and normal-dose CT (NDCT) image pairs to train denoising models, typically employing pixel-level loss functions like the mean-squared-error (MSE) loss. 
However, pixel-level supervision alone often leads to image over-smoothing, compromising subtle features such as texture. 
To mitigate this problem, Yang~\etal~\cite{wgan-vgg} combined the generative adversarial network (GAN) with a perceptual loss constraint, which effectively retains structural fidelity and improves visual impression.
Unfortunately, the inherent issues with GANs~\cite{huang2021gan,wgan-vgg} include unstable training, model collapse, and additional artifacts also known as hallucinations.

In this paper, we propose a novel \textbf{L}anguage-\textbf{E}ngaged \textbf{D}ual-space \textbf{A}lignment (\lossname) loss to offer an enhanced supervision over a LDCT denoising model, which aligns denoised CT with NDCT images in both the \textit{continuous perceptual space} and \textit{discrete semantic space} guided by a large language model (LLM). 
To this end, we first pretrain an LLM-guided CT autoencoder, which is built upon a vector quantized generative adversarial network (VQGAN)~\cite{vq-gan} that can encode a CT image into the continuous perceptual features and then quantized them into discrete vectors.
In contrast to the quantizer in VQGAN, we employ pretrained token embeddings from a frozen LLM as a substitute for the original learnable codebook~\cite{LQAE} and use a pyramid semantic loss in the Semantic Pyramid AutoEncoder (SPAE)~\cite{yu2024spae} to ensure that the quantized text tokens are semantically rich and aligned with anatomical information revealed by CT.
Subsequently, the \lossname loss is taken into account during the training process of the denoising model. By employing the LLM-guided CT autoencoder, \lossname enables minimization of the discrepancy between  denoised CT and NDCT images in terms of both the encoded continuous features and quantized discrete text embeddings.

By working in the continuous perceptual space and discrete semantic space, our denoising model with a minimized \lossname loss suppresses image noise, captures richer textures, and preserves finer details.
Moreover, our \lossname helps understand the anatomical semantic information in the denoised image with quantized text tokens during the denoising process, providing language-level explainability directly.

Our contributions are summarized as follows:
(1) We propose an \lossname loss to supervise LDCT denoising, which maximizes the similarity between the NDCT and denoised LDCT images in both the \textit{continuous perceptual space} and \textit{discrete semantic space} obtained using a LLM;
(2) Our \lossname synergizes the information in both the continuous and discrete space through an LLM-guided CT autoencoder, which is built on a VQGAN to quantize a CT image into semantic text tokens derived from the LLM’s vocabulary with a pyramid semantic loss; and
(3) Our extensive experimental results on two public LDCT denoising datasets demonstrate an encouraging image quality improvement with \lossname as 
a plug-and-play component offering language-level explainability.
To our best knowledge, this is the first work to apply LLM for LDCT denoising.

\begin{figure}[!ht]
\centering
\includegraphics[width=0.89\linewidth]{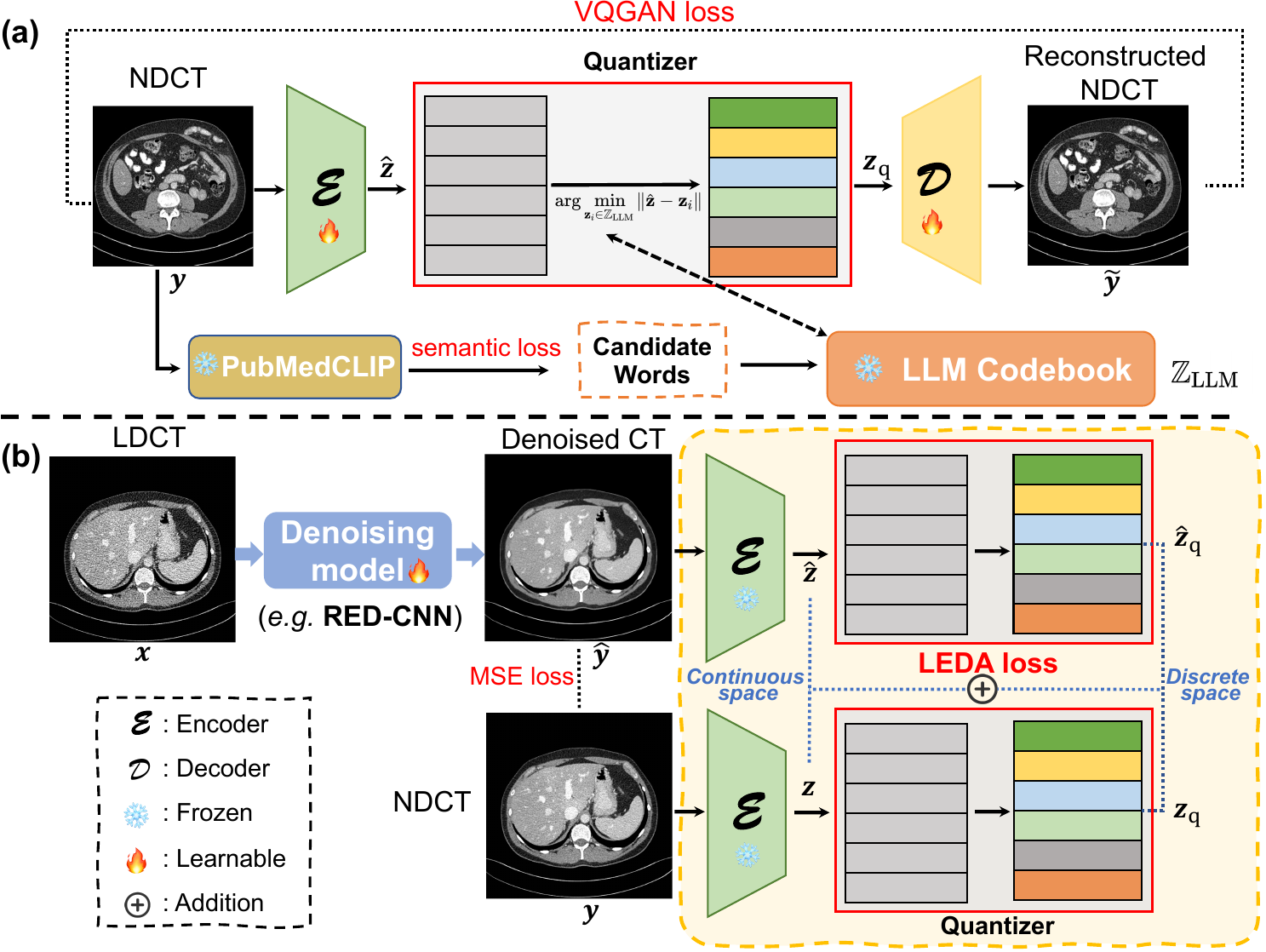}
\caption{Overview of the proposed \lossname. (a) Step 1: Training of the LLM-guided CT autoencoder; and (b) Step 2: Employment of \lossname for training  the denoising model.}
\label{network:IVTA}
\end{figure}

\section{Methodology}
In this section, we describe how to design the \lossname loss and use it to train a denoising model, as shown in 
Fig.~\ref{network:IVTA}.

\subsection{LLM-guided CT autoencoder}
\label{generator}
Our approach works in continuous feature space and discrete token embedding spaces through an LLM-guided CT autoencoder pretrained by NDCT images.
The autoencoder is built upon a VQ-GAN~\cite{vq-gan} architecture 
that consists of an encoder $\mathcal{E}$, decoder $\mathcal{D}$, and codebook $\mathbb{Z}$. Given an NDCT image $\mat{y} \in \mathbb{R}^{1 \times H \times W }$, $\mathcal{E}$ encodes it to produce high-level perceptual features $\mathcal{E}(\mat{y})$ = $\mat{\hat{z}} \in \mathbb{R}^{C \times H^{\prime} \times W^{\prime}}$ in the continuous space.
In the conventional vector quantizer, $\mat{\hat{z}}$ is then quantized into the discrete space through nearest neighbors lookup: $ \mat{z}_\mathrm{q} =\arg \min _{z_{i} \in \mathbb{Z}}\left\|\mat{\hat{z}}-\mat{z_{i}}\right\|$. However, these discrete representations lack concrete semantic information, which is very important for medical imaging tasks.

In order to address this semantic gap, we make two modifications on the original VQ-GAN quantizer by leveraging a pretrained LLM, as shown in Fig.~\ref{network:IVTA}(a). \textit{First}, we replace the learned codebook $\mathbb{Z}$  with a fixed codebook $\mathbb{Z_\mathrm{LLM}}=\{(t, \mat{e}(t)) \mid t \in \mathbb{V}\}$ from a pretrained LLM~\cite{LQAE}, where $\mathbb{V}$ is the LLM vocabulary and $\mat{e}(\cdot)$ produces the text embedding for a token $t$. 
It enables the encoder to map perceptual features into a discrete space delineated by text token embeddings, making the quantized representation directly convertible into text. 
However, these text tokens are random and may not be human understandable~\cite{LQAE}. 
\textit{Second}, we use a pyramid semantic loss from SPAE~\cite{yu2024spae} in the quantizer to learn hierachical semantic information. 
The quantized tokens consists of $D$ layers, with tokens at layer $l$ are denoted as  $\mat{t}_{l} \in \mathbb{V}^{h_{l} \times w_{l}}$. 
Concretely, we build per-layer candidate token pools $\mathbf{C}_{l}(\mat{y})=\{t \in \mathbb{V} \mid s(\mat{y}, t) \geq \rho_{l}\}$ according to the similarity 
score 
calculated by a large medical vision-language model PubMedCLIP~\cite{pubmedclip}, where $ s(\mat{y}, t)$ is the similarity score and $\rho_{l}$ is a threshold.
The pyramid semantic loss encourages the semantic similarity between an input NDCT image and each token in $\mathbf{C}_{l}(\mat{y})$, which is defined as
\begin{align}
    \mathcal{L}_{\text {semantic}}={\mathbb{E}}_{l \in\left[1, D\right]} {\mathbb{E}}_{\mathbf{z}_{l}} {\mathbb{E}}_{c \in \mathbf{C}_{l}(\mat{y})}-\log \frac{\exp(-\|(\mat{z}_{l}-\mat{e}(c) \|_{2}^{2})}{\sum_{k \in \mathbb{V}} \exp (-\|\mat{z}_{l}-\mat{e}(k)\|_{2}^{2})},
\end{align}
where $\mat{z}_{l}$ is the $l$-th layer embedding. 

The total training loss of the autoencoder encompasses both the VQGAN loss~\cite{vq-gan,LQAE} and the pyramid semantic loss: 
\begin{align}
&\mathcal{L}_{\text{total}}=\mathcal{L}_{\text{VQGAN}}+\alpha\omega\mathcal{L}_{\text {semantic}},
\end{align}
where $\mathcal{L}_{\text{VQGAN}}=\left\|\mat{y}-\tilde{\mat{y}}\right\|_{2}^{2}+\beta \sum_{l=1}^{D}\|\mat{z}-\operatorname{sg}(\hat{\mat{z}}_{\leq l})\|_{2}^{2}+\gamma \mathcal{L}_{\mathrm{GAN}}+\eta \mathcal{L}_{\text {Perceptual}}$, $\mathcal{L}_{\mathrm{GAN}}$ and $ \mathcal{L}_{\text {Perceptual}}$ are GAN~\cite{vq-gan} and perceptual~\cite{johnson2016perceptual} losses respectively, and $\hat{\mat{z}}_{\leq l}$ denotes the quantized token embeddings. Following~\cite{yu2024spae}, the hyperparameters $\alpha$, $\beta$, $\lambda$, 
 and $\eta$ are set to $0.3$, $0.3$, $0.1$, and $0.1$, respectively. $\omega$ is a dynamic weight of $(\frac{\mathcal{L}_{\text{VQGAN}}}{\mathcal{L}_{\text {semantic}}})$ without gradient backpropagation. Please refer to Fig.~\ref{supp_quantizer}  for more details on our quantizer.

\subsection{Employment of \lossname for training the denoising model}
\label{MSM}
After obtaining the LLM-guided CT autoencoder that can encode the CT image into continuous perceptual features and discrete text tokens, we are ready to define our \lossname loss, which employs the frozen autoencoder to align the denoised CT and NDCT images in both the continuous and discrete spaces, as shown in Fig.~\ref{network:IVTA}(b).
Given an LDCT image $\mat{x}$, we select a prevalent denoising model (\eg~RED-CNN~\cite{redcnn} or ESAU-Net~\cite{chen2023ascon}) to process it and obtain the denoised CT image $\mat{\hat{y}}$. 
Then, we use the pretrained LLM-guided CT autoencoder to encode both the denoised CT image $\mat{\hat{y}}$ and the ground-truth NDCT image $\mat{y}$, generating the continuous features $\mat{\hat{z}}$ and $\mat{z}$, respectively, which are then quantized  into text token embeddings $\mat{\hat{z}}_\mathrm{q}$ and $\mat{z}_\mathrm{q}.$

The \lossname loss is introduced to maximize the similarities 
between denoised CT and NDCT images by minimizing the Euclidean distance in terms of both continuous perceptual features and discrete semantic token embeddings.
During the training process, we combine the \lossname loss with the conventional pixel-level MSE loss to constrain the denoising model. The final denoising loss is defined as
\begin{equation}
\mathcal{L}=\mathcal{L}_\mathrm{MSE}+\lambda\mathcal{L}_\mathrm{\lossname}
=\|{\mat{y}}-\mat{\hat{y}}\|_2^{2}+\lambda(\underbrace{{\|\mat{z}}-\mat{\hat{z}}\|_2^{2}}_{\text{continuous}}+\underbrace{
\|{\mat{z}_\mathrm{q}}-\mat{\hat{z}}_\mathrm{q}\|_2^{2}}_{\text{discrete}}),
\end{equation}
where $\lambda$ is empirically set to $0.5$.

\section{Experiments}
\subsection{Experimental setup}
\noindent\textbf{Datasets.}\quad
We use two publicly available low-dose CT datasets: one released by the 2016 NIH AAPM-Mayo Clinic Low-Dose CT Grand Challenge~\cite{mccollough2017low} and a more recently released dataset~\cite{moen2021low}, which are referred to as Mayo-2016 and Mayo-2020 respectively. Mayo-2016 includes normal-dose abdominal CT images of 10 anonymous patients and corresponding simulated quarter-dose CT images. Mayo-2020 provides the abdominal CT image data of 100 patients with 25\% of normal dose, and we randomly choose 20 patients for our experiments.

\noindent\textbf{Implementation details.}\quad
For Mayo-2016, we selected 4800 image pairs of $512 \times 512$ images from 8 patients for training and 1136 pairs from 2 others for testing. 
For Mayo-2020, we used 2400 image pairs of $512 \times 512$ images from 16 patients for training and 580 pairs from the remaining 4 patients for testing.
Our LLM codebook is taken from the token embedding layer from PubMedCLIP~\cite{pubmedclip}.
Both the encoder and decoder of VQGAN are composed of 6 ResNet blocks and 2 attention blocks~\cite{vq-gan}. 
We apply the pyramid semantic loss with a 3-layer token pyramid (see in Fig.~\ref{supp_quantizer}, with 4, 64 and 1024 tokens per layer and thresholds of 0.95, 0.9 and 0.8 respectively.
We use an NVIDIA GeForce RTX 3090 to train all methods with a  window of $[-1000, 2000]$ Hounsfield unit (HU).
For the autoencoder, we follow the setting of Ref.\cite{vq-gan} with batch size of 4. 
All denoising models are optimized with AdamW~\cite{loshchilov2017decoupled} ($\beta_{1}=0.9$, $\beta_{2}=0.99$, weight decay $1.0\times10^{-9}$), with the initial learning rate of $1.0\times 10^{-4}$, decreasing to $1.0\times 10^{-6}$ via cosine annealing~\cite{loshchilov2016sgdr}.

\begin{table}[b]
\caption{Performance comparison on the Mayo-2016 dataset.}
\label{Results_comparison_1}
\centering
\begin{tabular*}{1\linewidth}{@{\extracolsep{\fill}}rcccc}
\shline
\textbf{Methods} && PSNR$\uparrow$ &  SSIM$\uparrow$   &   FSIM$\uparrow$ \\
\midrule

{RED-CNN~\cite{redcnn}}   && 28.69\std{$\pm${1.57}} & 0.8562\std{$\pm${0.0381}} & 0.9321\std{$\pm${0.0180}} \\   

{EASU-Net~\cite{chen2023ascon}}   && \textbf{28.74}\std{$\pm${1.58}} & 0.8616\std{$\pm${0.0386}} & 0.9353\std{$\pm${0.0188}}   \\

{WGAN-VGG}~\cite{wgan-vgg}  && 27.09\std{$\pm${1.61}}  &  0.8558\std{$\pm${0.0389}}  & 0.9390\std{$\pm${0.0189}}  \\

{DU-GAN~\cite{huang2021gan}}  && 27.51\std{$\pm${1.58}}  & 0.8572\std{$\pm${0.0371}}  &  0.9407\std{$\pm${0.0136}}    \\

\hline

{RED-CNN+$\mathcal{L}_\mathrm{perceptual}$} &&   27.47\std{$\pm${1.68}} & 0.8542\std{$\pm${0.0385}} &   0.9363\std{$\pm${0.0155}}   \\

{EASU-Net+$\mathcal{L}_\mathrm{perceptual}$}   && 27.58\std{$\pm${1.82}} & 0.8573\std{$\pm${0.0387}} & 0.9385\std{$\pm${0.0161}}    \\

\hline

{RED-CNN+$\mathcal{L}_\mathrm{\lossname}$} && 28.59\std{$\pm${1.47}} & 0.8636\std{$\pm${0.0357}} & 0.9393\std{$\pm${0.0175}}   \\

{EASU-Net+$\mathcal{L}_\mathrm{\lossname}$}   && 28.64\std{$\pm${1.59}} & {\textbf{0.8647}}\std{$\pm${0.0373}} & \textbf{0.9410}\std{$\pm${0.0161}}    \\ 
\shline
\end{tabular*}
\end{table}

\begin{table}[t]
\caption{Performance comparison on the Mayo-2020 dataset.}
\label{Results_comparison_2}
\centering
\begin{tabular*}{1\linewidth}{@{\extracolsep{\fill}}rcccc}
\shline
\textbf{Methods} && PSNR$\uparrow$ &  SSIM$\uparrow$   & FSIM$\uparrow$ \\
\midrule

{RED-CNN~\cite{redcnn}}   && 34.03\std{$\pm${1.80}}  & 0.9328\std{$\pm${0.0217}} & 0.9686\std{$\pm${0.0101}} 
\\

{ESAU-Net~\cite{chen2023ascon}}   && \textbf{34.12}\std{$\pm${1.93}}  & 0.9334\std{$\pm${0.0218}} & 0.9708\std{$\pm${0.0100}} \\

{WGAN-VGG}~\cite{wgan-vgg}  && 32.82\std{$\pm${1.74}}  & 0.9245\std{$\pm${0.0252}}  & 0.9698\std{$\pm${0.0101}}  
\\

{DU-GAN~\cite{huang2021gan}}  && 32.55\std{$\pm${1.36}}  & 0.9203\std{$\pm${0.0270}}  &  0.9654\std{$\pm${0.0107}}  
\\

\hline

{RED-CNN+$\mathcal{L}_\mathrm{perceptual}$} && 33.43\std{$\pm${1.73}} &0.9310\std{$\pm${0.0229}}  &  0.9706\std{$\pm${0.0097}}  
\\

{ESAU-Net+$\mathcal{L}_\mathrm{perceptual}$}  && 33.49\std{$\pm${1.93}}  & 0.9328\std{$\pm${0.0228}} &  0.9720\std{$\pm${0.0095}}  \\

\hline

{RED-CNN+$\mathcal{L}_\mathrm{\lossname}$} &&  33.97\std{$\pm${1.75}} &0.9343\std{$\pm${0.0216}}  &  0.9721\std{$\pm${0.0096}}  
\\

{ESAU-Net+$\mathcal{L}_\mathrm{\lossname}$}  && 34.06\std{$\pm${1.92}}  & \textbf{0.9351}\std{$\pm${0.0220}} & \textbf{0.9731}\std{$\pm${0.0093}} 
\\
\shline
\end{tabular*}
\end{table}

\begin{figure}[!h]
\centering
\includegraphics[width=0.9\linewidth]{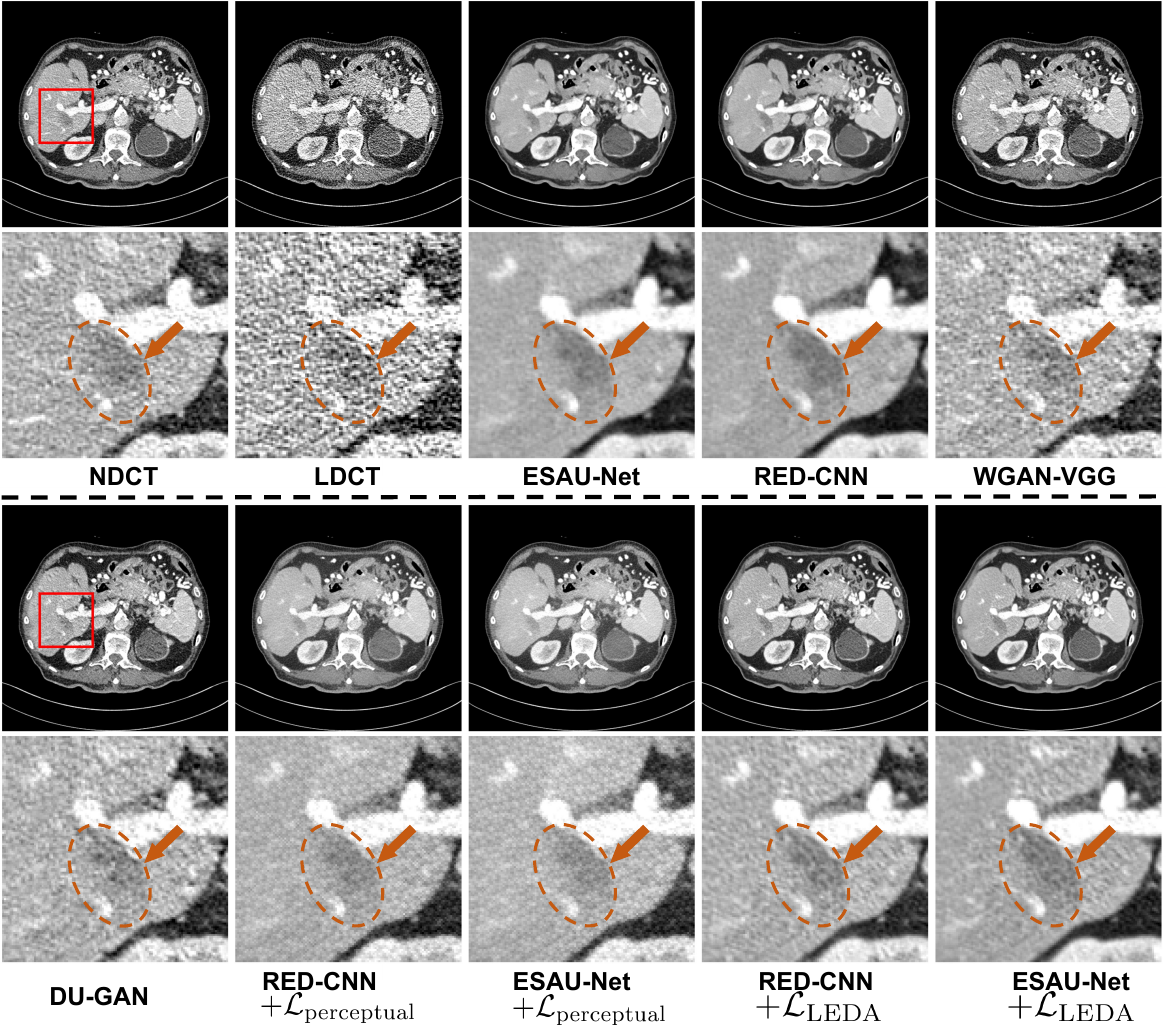}
\caption{Transverse CT images from the Mayo-2016 dataset. The ROI of the rectangle is zoomed below  for better visualization. The display window is [-160, 240] HU.
}
\label{test_results_spatial}
\end{figure}

\subsection{Performance comparison}
\noindent\textbf{Quantitative evaluation.}\quad
We use three widely used metrics including peak signal-to-noise ratio (PSNR), structural similarity index (SSIM), and feature similarity index (FSIM). 
Since both datasets we used are abdominal data, we compute the quantitative metrics on the abdominal window of $[-160,240]$ HU.
Tables~\ref{Results_comparison_1} and~\ref{Results_comparison_2} present the testing results on the Mayo-2016 and Mayo-2020 datasets, respectively. 
We compare our \lossname loss with perceptual loss~\cite{johnson2016perceptual} and two GAN loss based methods~(\ie~WGAN-VGG~\cite{wgan-vgg} and DU-GAN~\cite{huang2021gan}). The generator of both WGAN-VGG and DU-GAN is RED-CNN.
Tables~\ref{Results_comparison_1} and~\ref{Results_comparison_2} show that although RED-CNN and ESAU-Net with solely pixel-level supervision achieve better PSNR results, the visual inspection of Fig.~\ref{test_results_spatial} reveals over-smoothed features as compared to the NDCT images, indicating a loss of structural information.
Adding the \lossname loss, RED-CNN and ESAU-Net perform the best in terms of SSIM and FSIM scores on both datasets  with high visual fidelity while PSNR is also better than GAN-base methods.

\noindent\textbf{Qualitative evaluation.}\quad
Fig.~\ref{test_results_spatial} presents a representative qualitative result on Mayo-2016, where the area of interest is indicated by the arrow.
It can be observed that RED-CNN and ESAU-Net greatly reduce the noise but cause the over-soothing problem.
Although the GAN-based methods alleviate this problem and retain texture more faithfully, they suffer from incomplete denoising and introduce velvet artifacts, likely due to the unstable training process. 
In addition, while the utilization of the perceptual loss  enhances textural information, it blurs feature details and compromises image fidelity to some degree. 
In contrast to the aforementioned approaches, our \lossname loss excels in not only reducing image noise but also preserving structural details without introducing artifacts. This benefits from the synergistic combination of continuous perceptual features and discrete semantic tokens.
Please refer to Fig.~\ref{supp_spatial} for more qualitative results on Mayo-2020.

\begin{figure}[t]
\centering
\includegraphics[width=0.9\linewidth]{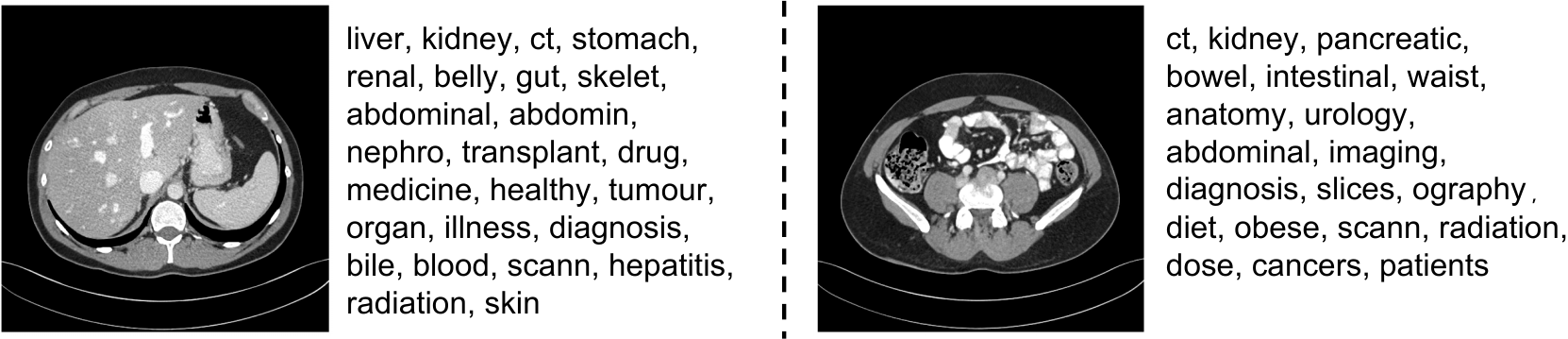}
\caption{Explainability provided by the text tokens extracted from the LLM, in which only part of quantized tokens are shown for the first 2 layers in the quantizer.}
\label{interpretability}
\end{figure}

\noindent\textbf{Text token-aided explainability.}\quad
In addition to the performance improvement, Fig.~\ref{interpretability} shows that our \lossname can provide language-level explainability from quantized tokens.
Specifically, we select quantized token embeddings from the first two layers of the token pyramid and identify the corresponding tokens from the LLM vocabulary.
Note that the semantic information decoded by the quantized token embeddings exhibits a certain similarity with the image content. 
This includes not only global concepts, such as `dose', `ct', and `abdominal', but also more granular details, like `organ', `cancer', and `liver'. 
To our knowledge, this represents an initial effort to leverage the LLM for LDCT denoising, enhancing image quality and providing explainability at the same time. 

\begin{table*}[htbp]
\caption{Ablation results on the different components in \lossname with RED-CNN as the denoising backbone.}
\label{table:loss_abla}
\renewcommand\arraystretch{0.95}
\begin{tabular*}{1\linewidth}{@{\extracolsep{\fill}}cccccc}
\shline
 &\multicolumn{2}{c}{\lossname loss} & 
 \multirow{2}{*}{PSNR$\uparrow$} & \multirow{2}{*}{SSIM$\uparrow$} & \multirow{2}{*}{FSIM$\uparrow$} \\
 \cline{2-3}
 & \lossname-C         & \lossname-D               &      &      \\ 
 \midrule
 &   \checkmark  &    \textbf{-}   &  28.34\std{$\pm${1.58}}  & \textbf{0.8640}\std{$\pm${0.0353}}  &  \textbf{0.9404}\std{$\pm${0.0134}}  
 \\
 &  \textbf{-}     &   \checkmark       &   \textbf{28.70}\std{$\pm${1.57}}    &  0.8615\std{$\pm${0.0371}} & 0.9360\std{$\pm${0.0161}} 
 \\
 &   \checkmark     &    \checkmark  & \underline{28.59}\std{$\pm${1.47}} & \underline{0.8636}\std{$\pm${0.0357}} & \underline{0.9393}\std{$\pm${0.0175}}    \\
\shline
\end{tabular*}
\end{table*}

\begin{figure}[h]
\centering
\includegraphics[width=0.9\linewidth]{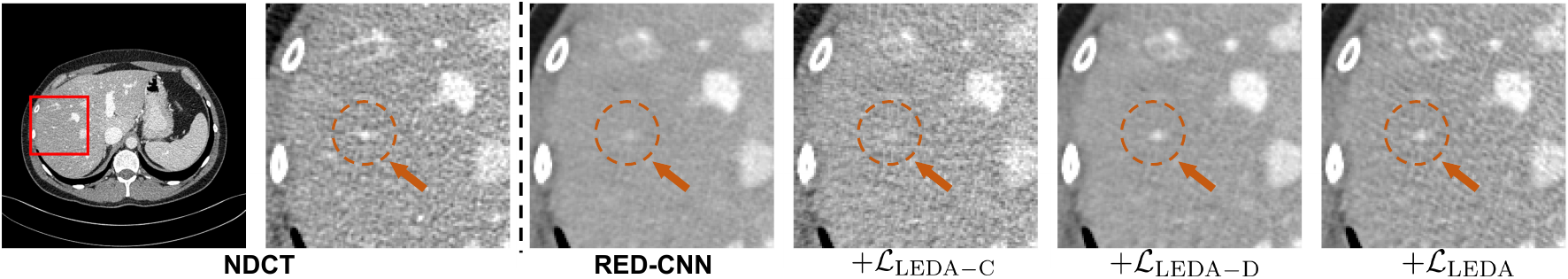}
\caption{Transverse CT images of ablation studies on the different components in \lossname.}
\label{fig:loss_abla}
\end{figure}

\noindent\textbf{Ablation studies.}\quad
To demonstrate the effectiveness of the proposed \lossname. We start with RED-CNN~\cite{redcnn} using MSE and gradually insert two components in our \lossname loss. We use \lossname-C and \lossname-D to denote the loss with only continuous space alignment and only discrete space alignment, respectively. Table~\ref{table:loss_abla} and Fig.~\ref{fig:loss_abla} present the results, which show that the alignment of the continuous perceptual space retains richer texture and simultaneously obtains better SSIM and FSIM scores. On the other hand, \lossname-D, focusing on the discrete semantic space alignment, preserves finer details, thereby elevating the PSNR metric. Our \lossname achieves an optimal balance through the alignment of the two spaces and obtains the results closest to the NDCT. 

The results in Table~\ref{limitation} verify the importance of autoencoder's reconstruction quality.
First, using an LLM codebook negatively affects the reconstruction results, but with semantic guidance it performs comparably with the original VQGAN while producing semantic text tokens.
Second, the reconstruction quality of the LLM-guided CT autoencoder directly affects the performance of LDCT denoising.
We believe that improving the NDCT reconstruction quality could further elevate the performance of the \lossname loss.

\begin{table}[t]
\centering
\caption{Reconstruction and denoising results of \lossname constructed by different types of VQGAN on Mayo-2016 with RED-CNN as the denoising backbone.}
\label{limitation}
\begin{tabular*}{1\linewidth}{@{\extracolsep{\fill}}lccc}
\shline
   & \textbf{Reconstruction quality} & \multicolumn{2}{c}{\textbf{Denoising quality}}  \\
    &    FID   & PSNR & SSIM \\
 \hline
VQGAN & 58.17& 28.55 &  0.8611 \\
\quad\quad+LLM codebook & 62.18& 28.49& 0.8571 \\
\quad\quad\quad\quad+semantic loss & \textbf{51.79}& \textbf{28.59} & \textbf{0.8636}\\
\shline
\end{tabular*}
\end{table}

\section{Conclusion}
In this paper, we have proposed the \lossname loss that empowers the LDCT denoising model performance by aligning the denoised image with the NDCT image in both the {continuous perceptual space} and {discrete semantic space}. 
We have synergized the information in these two spaces using an LLM-guided CT autoencoder, which encodes a CT image into continuous high-level features and quantizes them into text tokens derived from the LLM’s vocabulary.
Extensive results have validated the performance improvement of the proposed alignments. This is the first effort to apply LLM for LDCT denoising, simultaneously enhancing denoising quality and providing language-level explainability.

We acknowledge some limitations in this work. First, the reconstruction quality of our LLM-guided CT autoencoder still falls short compared to the state-of-the-art models in image generation, which is expected to be improved in the future to further improve denoising performance.
Second, we only exhibit tokens from the first 2 layers of the token pyramid in Fig.~\ref{interpretability} due to the poor correlation between many tokens in the 3-th layer and CT anatomical information. 
Finer alignment of the image with tokens is worthy of studying in future.

\clearpage

\section*{Supplementary Materials}
\renewcommand*{\thefigure}{S\arabic{figure}}
\setcounter{figure}{0}
\renewcommand*{\thetable}{S\arabic{table}}
\setcounter{table}{0}

\begin{figure}[h]
\centering
\includegraphics[width=.7\linewidth]
{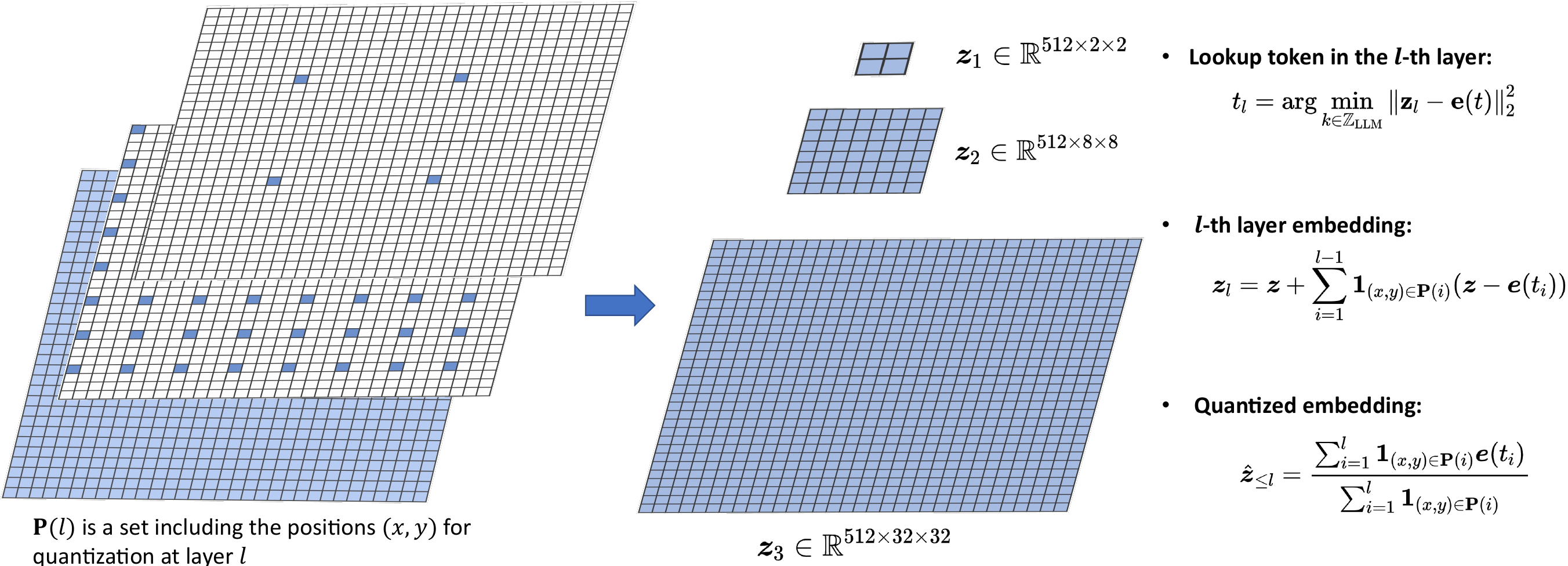}
\caption{The 3-layer token pyramid in our LLM-guided CT autoencoder. We select position in using a downsample ratio of 4.}
\label{supp_quantizer}
\end{figure}
\begin{figure}[htbp]
\centering
\includegraphics[width=.6\linewidth]{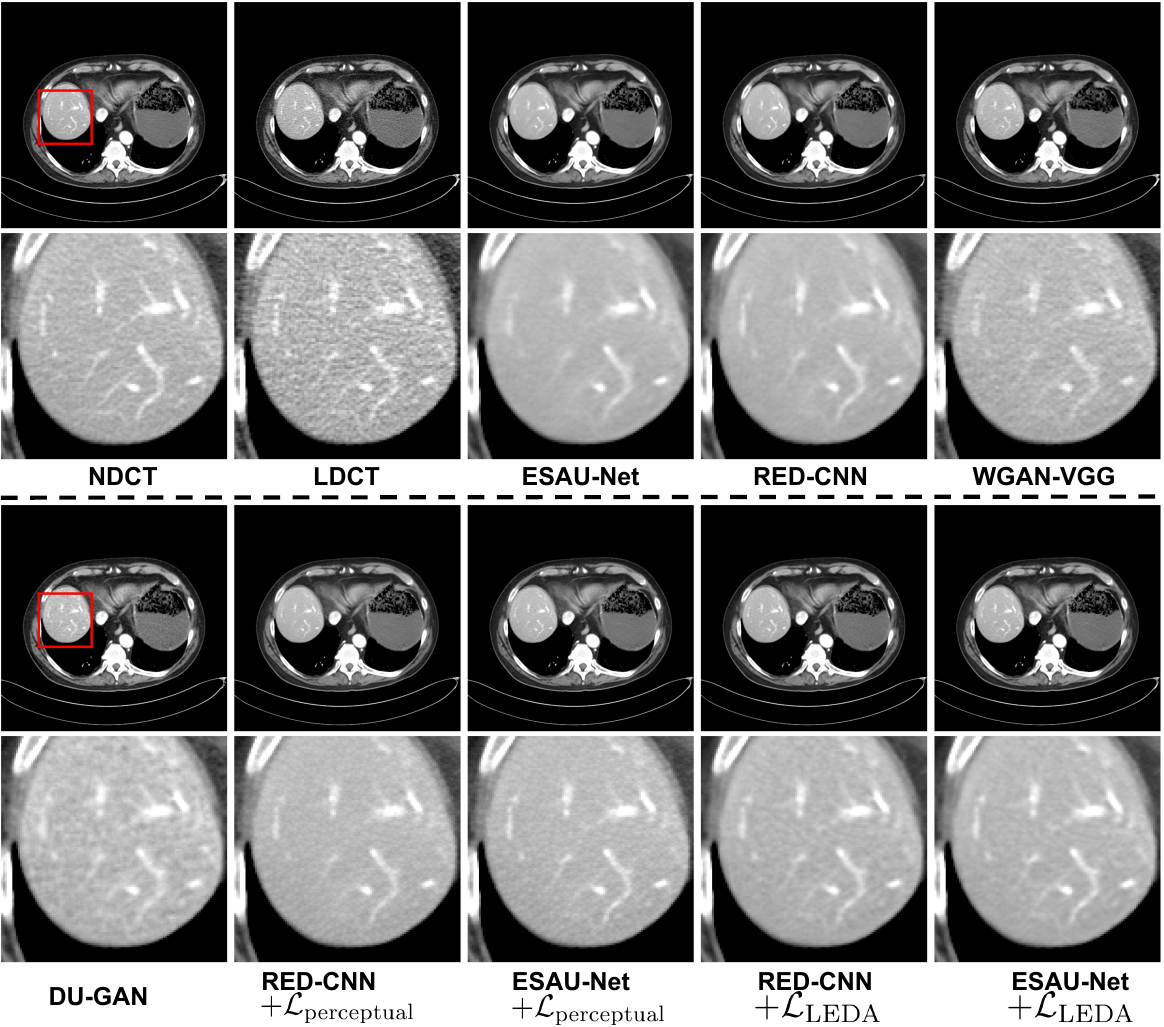}
\caption{Transverse CT images and corresponding ROI images from the Mayo-2020 dataset. It can be observed that our LEDA loss excels in not only reducing image noise but also preserving structural details without introducing artifacts.}
\label{supp_spatial}
\end{figure}


\begin{thebibliography}{10}
\providecommand{\url}[1]{\texttt{#1}}
\providecommand{\urlprefix}{URL }
\providecommand{\doi}[1]{https://doi.org/#1}

\bibitem{redcnn}
Chen, H., Zhang, Y., Kalra, M.K., Lin, F., Chen, Y., Liao, P., Zhou, J., Wang, G.: Low-dose {CT} with a residual encoder-decoder convolutional neural network. IEEE Trans. Med. Imaging  \textbf{36}(12),  2524--2535 (2017)

\bibitem{chen2023ascon}
Chen, Z., Gao, Q., Zhang, Y., Shan, H.: {ASCON}: Anatomy-aware supervised contrastive learning framework for low-dose {CT} denoising. In: MICCAI. pp. 355--365. Springer (2023)

\bibitem{litformer}
Chen, Z., Niu, C., Gao, Q., Wang, G., Shan, H.: {LIT-Former}: Linking in-plane and through-plane transformers for simultaneous {CT} image denoising and deblurring. IEEE Trans. Med. Imaging  (2024)

\bibitem{pubmedclip}
Eslami, S., de~Melo, G., Meinel, C.: Does {CLIP} benefit visual question answering in the medical domain as much as it does in the general domain? arXiv preprint arXiv:2112.13906  (2021)

\bibitem{vq-gan}
Esser, P., Rombach, R., Ommer, B.: Taming transformers for high-resolution image synthesis. In: Proc. IEEE Conf. Comput. Vis. Pattern Recognit. pp. 12873--12883 (2021)

\bibitem{corediff}
Gao, Q., Li, Z., Zhang, J., Zhang, Y., Shan, H.: {CoreDiff}: Contextual error-modulated generalized diffusion model for low-dose {CT} denoising and generalization. IEEE Trans. Med. Imaging  \textbf{43}(2),  745--759 (2024)

\bibitem{ge2022ddpnet}
Ge, R., He, Y., Xia, C., Sun, H., Zhang, Y., Hu, D., Chen, S., Chen, Y., Li, S., Zhang, D.: {DDPNet}: a novel dual-domain parallel network for low-dose {CT} reconstruction. In: MICCAI. pp. 748--757. Springer (2022)

\bibitem{ge2023jccs}
Ge, R., He, Y., Xia, C., Zhang, D.: {JCCS-PFGM}: A novel circle-supervision based poisson flow generative model for multiphase {CECT} progressive low-dose reconstruction with joint condition. In: MICCAI. pp. 409--418. Springer (2023)

\bibitem{gholizadeh2020deep}
Gholizadeh-Ansari, M., Alirezaie, J., Babyn, P.: Deep learning for low-dose {CT} denoising using perceptual loss and edge detection layer. J. Digit. Imaging  \textbf{33},  504--515 (2020)

\bibitem{huang2021gan}
Huang, Z., Zhang, J., Zhang, Y., Shan, H.: {DU-GAN}: Generative adversarial networks with dual-domain {U-Net-based} discriminators for low-dose {CT} denoising. IEEE Trans. Instrum. Meas.  \textbf{71},  1--12 (2021)

\bibitem{jing2023inter}
Jing, J., Wang, T., Yu, H., Lu, Z., Zhang, Y.: Inter-slice consistency for unpaired low-dose {CT} denoising using boosted contrastive learning. In: MICCAI. pp. 238--247. Springer (2023)

\bibitem{johnson2016perceptual}
Johnson, J., Alahi, A., Fei-Fei, L.: Perceptual losses for real-time style transfer and super-resolution. In: Proc. European Conf. Comput. Vis. pp. 694--711. Springer (2016)

\bibitem{lei2023ct}
Lei, Y., Niu, C., Zhang, J., Wang, G., Shan, H.: {CT} image denoising and deblurring with deep learning: Current status and perspectives. IEEE Trans. Radiat. Plasma Med. Sci. \textbf{8}(2), 153-172, (2024)

\bibitem{LQAE}
Liu, H., Yan, W., Abbeel, P.: Language quantized autoencoders: Towards unsupervised text-image alignment. Proc. Adv. Neural Inf. Process.  \textbf{36} (2024)

\bibitem{loshchilov2016sgdr}
Loshchilov, I., Hutter, F.: {SGDR}: Stochastic gradient descent with warm restarts. arXiv preprint arXiv:1608.03983  (2016)

\bibitem{loshchilov2017decoupled}
Loshchilov, I., Hutter, F.: Decoupled weight decay regularization. arXiv preprint arXiv:1711.05101  (2017)

\bibitem{mccollough2017low}
McCollough, C.H., Bartley, A.C., Carter, R.E., Chen, B., Drees, T.A., Edwards, P., Holmes~III, D.R., Huang, A.E., Khan, F., Leng, S., et~al.: Low-dose {CT} for the detection and classification of metastatic liver lesions: results of the 2016 low dose {CT} grand challenge. Med. Phys.  \textbf{44}(10),  e339--e352 (2017)

\bibitem{moen2021low}
Moen, T.R., Chen, B., Holmes~III, D.R., Duan, X., Yu, Z., Yu, L., Leng, S., Fletcher, J.G., McCollough, C.H.: Low-dose {CT} image and projection dataset. Med. Phys.  \textbf{48}(2),  902--911 (2021)

\bibitem{niu2022noise}
Niu, C., Li, M., Fan, F., Wu, W., Guo, X., Lyu, Q., Wang, G.: Noise suppression with similarity-based self-supervised deep learning. IEEE Trans. Med. Imaging  (2022)

\bibitem{shan2019competitive}
Shan, H., Padole, A., Homayounieh, F., Kruger, U., Khera, R.D., Nitiwarangkul, C., Kalra, M.K., Wang, G.: Competitive performance of a modularized deep neural network compared to commercial algorithms for low-dose {CT} image reconstruction. Nat. Mach. Intell.  \textbf{1}(6),  269--276 (2019)

\bibitem{shan20183}
Shan, H., Zhang, Y., Yang, Q., Kruger, U., Kalra, M.K., Sun, L., Cong, W., Wang, G.: {3-D} convolutional encoder-decoder network for low-dose {CT} via transfer learning from a {2-D} trained network. IEEE Trans. Med. Imaging  \textbf{37}(6),  1522--1534 (2018)

\bibitem{shen2022mlf}
Shen, J., Luo, M., Liu, H., Liao, P., Chen, H., Zhang, Y.: {MLF-IOSC}: Multi-level fusion network with independent operation search cell for low-dose {CT} denoising. IEEE Trans. Med. Imaging  \textbf{42}(4),  1145--1158 (2022)

\bibitem{wang2023ctformer}
Wang, D., Fan, F., Wu, Z., Liu, R., Wang, F., Yu, H.: {CTformer}: convolution-free token2token dilated vision transformer for low-dose {CT} denoising. Phys. Med. Biol.  \textbf{68}(6),  065012 (2023)

\bibitem{wang2023masked}
Wang, D., Xu, Y., Han, S., Yu, H.: Masked autoencoders for low-dose {CT} denoising. In: Proc. IEEE Int. Symp. Biomed. Imaging. pp.~1--4. IEEE (2023)

\bibitem{wang2018image}
Wang, G., Ye, J.C., Mueller, K., Fessler, J.A.: Image reconstruction is a new frontier of machine learning. IEEE Trans. Med. Imag.  \textbf{37}(6),  1289--1296 (2018)

\bibitem{wgan-vgg}
Yang, Q., Yan, P., Zhang, Y., Yu, H., Shi, Y., Mou, X., Kalra, M.K., Zhang, Y., Sun, L., Wang, G.: Low-dose {CT} image denoising using a generative adversarial network with wasserstein distance and perceptual loss. IEEE Trans. Med. Imaging  \textbf{37}(6),  1348--1357 (2018)

\bibitem{yu2024spae}
Yu, L., Cheng, Y., Wang, Z., Kumar, V., Macherey, W., Huang, Y., Ross, D., Essa, I., Bisk, Y., Yang, M.H., et~al.: {SPAE}: Semantic pyramid autoencoder for multimodal generation with frozen {LLMs}. Proc. Adv. Neural Inf. Process.  \textbf{36} (2024)

\end{thebibliography}
\end{document}